\documentstyle[prl,aps,psfig]{revtex}
\newcommand{\beq}{\begin{equation}}
\newcommand{\eeq}{\end{equation}}
\newcommand{\beqa}{\begin{eqnarray}}
\newcommand{\eeqa}{\end{eqnarray}}
\newcommand{\ba}{\begin{array}}
\newcommand{\ea}{\end{array}}

\begin{document}
\draft

\twocolumn[\hsize\textwidth\columnwidth\hsize\csname
@twocolumnfalse\endcsname

\widetext 
\title{Bose-condensed Bright Solitons under Transverse Confinement} 
\author{L. Salasnich$^{1}$, A. Parola$^{2}$ and L. Reatto$^{1}$} 
\address{$^{1}$Istituto Nazionale per la Fisica della Materia, 
Unit\`a di Milano \\
Dipartimento di Fisica, Universit\`a di Milano, \\
Via Celoria 16, 20133 Milano, Italy\\
$^{1}$Istituto Nazionale per la Fisica della Materia, 
Unit\`a di Como \\
Dipartimento di Scienze Fisiche, Universit\`a dell'Insubria, \\
Via Valeggio 11, 23100 Como, Italy}
\maketitle

\begin{abstract} 
We investigate the dynamics of Bose-condensed bright solitons made of 
alkali-metal atoms with negative scattering length and under harmonic 
confinement in the transverse direction. 
Contrary to the 1D case, the 3D bright soliton 
exists only below a critical attractive interaction which depends 
on the extent of confinement. Such a behavior is also found in multi-soliton 
condensates with box boundary conditions. 
We obtain numerical and analytical estimates of the critical 
strength beyond which the solitons do not exist. 
By using an effective 1D nonpolynomial 
nonlinear Schr\"odinger equation (NPSE), which 
accurately takes into account the transverse dynamics 
of cigar-like condensates, we numerically simulate the dynamics 
of the "soliton train" reported in a recent experiment 
(Nature {\bf 417} 150 (2002)). Then, 
analyzing the macroscopic quantum tunneling of the bright soliton 
on a Gaussian barrier we find that its interference in the tunneling region 
is strongly suppressed with respect to non-solitonic case;  
moreover, the tunneling through a barrier breaks the 
solitonic nature of the matter wave. 
Finally, we show that the collapse of the soliton is induced 
by the scattering on 
the barrier or by the collision with another matter wave when 
the density reaches a critical value, for which we derive an accurate 
analytical formula. 
\end{abstract}


]

\narrowtext


\section{Introduction} 

The experimental achievement of Bose-Einstein condensation 
with alkali-metal atoms at ultra-low temperatures [1,2] 
has opened the exciting possibility of studying 
topological configurations of the Bose-Einstein condensate (BEC), 
like solitary waves (solitons) with positive or negative 
scattering length $a_s$ [3]. Dark solitons ($a_s>0$) 
of Bose condensed atoms have been experimentally observed 
few years ago [4], while bright solitons ($a_s<0$) 
have been detected only very recently with $^7$Li using an 
optical red-detuned laser beam along the axial direction of the 
sample to impose a transverse (radial) confinement [5]. 
\par 
Bright solitons have been studied in one-dimensional (1D) models 
[6] and no investigation has been performed on the 
effects a finite transverse width which is always present in experiment. 
In this paper the dynamics of a Bose condensate is investigated 
by using the full 3D Gross-Pitaevskii equation [7] and 
an effective 1D nonpolynomial 
nonlinear Schr\"odinger equation (NPSE) [8,9], which 
accurately takes into account the transverse dynamics 
for cigar-like condensates. We analyze the existence, 
stability and collective oscillations 
of these 3D Bose-condensed bright solitons 
starting from a BEC with transverse confinement. 
We also study the existence and stability of multi-soliton 
condensates in a box and simulate the dynamics of 
the "soliton train" experimentally observed in Ref. [5]. 
Then we investigate the macroscopic quantum 
tunneling of a BEC on a Gaussian barrier showing that 
the bright-soliton strongly reduces interference fringes in the 
tunneling region. We show that the collapse of 
BEC can be induced by its scattering on a barrier 
if its density reaches a critical value, 
which can be analytically predicted. Finally we study the scattering 
of bright solitons and the conditions for their collapse at the collision. 

\section{Bright solitons under transverse confinement} 

The dynamics of a BEC at zero temperature is well described by 
the 3D Gross-Pitaevskii equation (3D GPE) [7]. In many cases the numerical 
solution of 3D GPE is a hard task and approximate procedures are needed. 
Starting from the 3D GPE we have recently derived and studied [8,9] 
an effective time-dependent 
1D nonpolynomial nonlinear Schr\"odinger equation (NPSE). 
NPSE describes very accurately Bose condensates 
confined by a harmonic potential with frequency $\omega_{\bot}$ 
and harmonic length $a_{\bot}=(\hbar/m\omega_{\bot})^{1/2}$ 
in the transverse direction and by a generic potential $V(z)$ 
in the axial one. The total wave function of the condensate is 
$\psi(x,y,z,t) = f(z,t) \phi(x,y,t)$, where the transverse 
wavefunction $\phi(x,y,t)$ is a Gaussian with a width $\eta$ 
given by $\eta^2=a_{\bot}^2\sqrt{1+2a_sN|f|^2}$ and 
$f(z,t)$ satisfies the NPSE equation 
$$
i\hbar {\partial \over \partial t}f= 
\left[ -{\hbar^2\over 2m} {\partial^2\over \partial z^2} 
+ V(z) + {2 \hbar^2 N a_s \over m a_{\bot}^2} 
{|f|^2\over \sqrt{1+ 2a_s N|f|^2} } 
\right. 
$$ 
\beq 
\left. 
+ {\hbar \omega_{\bot}\over 2}  
\left( {1\over \sqrt{1+ 2 a_s N|f|^2} } + \sqrt{1+ 2a_s N|f|^2}
\right) \right] f  \; , 
\eeq 
where $a_s$ is the s-wave scattering length. 
$N$ is the number of condensed 
bosons and the function $f(z,t)$ is normalized to one. 
In the weakly-interacting limit $a_s N|f|^2 <<1$, 
NPSE reduces to a 1D GPE. Instead, in the strongly-interacting 
limit, NPSE becomes a nonlinear Sch\"odinger equation with 
the nonlinear term proportional to $|f|f$ [8,9]. 
\par 
Under transverse confinement and negative scattering 
length ($a_s<0$) a bright soliton sets up when the negative 
inter-atomic energy of the BEC compensates the positive kinetic energy 
such that the BEC is self-trapped in the axial direction. 
The shape of this 3D Bose-condensed bright soliton 
can be deduced from NPSE [8]. Setting $V(z)=0$, 
scaling $z$ in units of $a_{\bot}$ and $t$ in units 
of $\omega_{\bot}^{-1}$, with the position 
\beq 
f(z,t)=\Phi(z-vt) e^{iv(z-vt)} e^{i(v^2/2 - \mu)t} \; , 
\eeq 
one finds the bright-soliton 
solution written in implicit form 
$$ 
\zeta= {1\over \sqrt{2}} {1\over \sqrt{1-\mu}} \; 
arctg\left[ 
\sqrt{ \sqrt{1-2\gamma\Phi^2}-\mu \over 1-\mu } 
\right] 
$$
\beq
-{1\over \sqrt{2}} {1\over \sqrt{1+\mu}} \; 
arcth\left[ 
\sqrt{ \sqrt{1-2\gamma\Phi^2}-\mu \over 1+\mu } 
\right] \; ,  
\eeq 
where $\zeta =z-vt$ and $\gamma=|a_s|N/a_{\bot}$. 
This equation is well defined only for $\gamma \Phi^2 <1/2$; 
at $\gamma \Phi^2 = 1/2$ the transverse size is zero. 
Moreover, by imposing the normalization condition one has 
\beq 
(1-\mu)^{3/2} - {3\over2} (1-\mu)^{1/2} + 
{3\over 2 \sqrt{2}} \gamma = 0 \; . 
\eeq 
The normalization relates the chemical potential $\mu$ 
to the coupling constant $\gamma$, 
while the velocity $v$ of the bright soliton 
remains arbitrary. 
In the weak-coupling limit ($\gamma\Phi^2<<1$), 
the normalization condition gives 
$\mu= 1 - \gamma^2/2$ and the bright-soliton solution reads 
\beq
\Phi(\zeta)=\sqrt{\gamma\over 2} \; sech\left[{\gamma}\zeta \right] \; .  
\eeq 
The above solution is the text-book 1D bright soliton 
of the 1D nonlinear (cubic) Schr\"odinger equation (1D GPE). 
\par
In Figure 1 the ``exact'' solitonic solution obtained 
by numerically solving the 3D GPE with a finite-difference 
predictor-corrector method [10] is compared with 
the analytical solution (Eq. 3) of the NPSE and the analytical solution 
(Eq. 5) of the 1D bright soliton. The density profile is plotted for 
increasing values of $\gamma$. The agreement between the 
``exact'' solution and the analytical one obtained from the NPSE is 
always remarkably good. Instead, as expected, the profile of the 1D bright 
soliton deviates from the 3D one for large $\gamma$. 

\begin{figure}
\centerline{\psfig{file=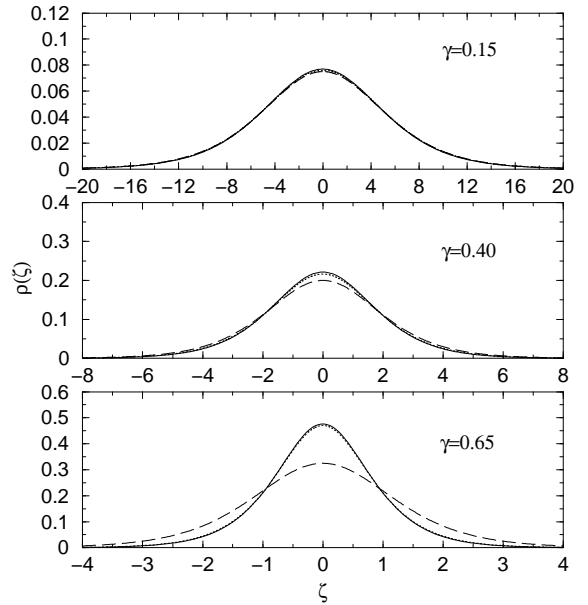,height=3.2in}}
\caption{Axial density profile $\rho(z)$ of the Bose-condensed 
bright soliton: 3D GPE (full line), Eq. 3 (dotted line), 
Eq. 5 (dashed line). 
Length in units $a_{\bot}=(\hbar /m\omega_{\bot})^{1/2}$ 
and density in units $1/a_{\bot}$. } 
\end{figure}

\par 
From Eq. 4 it easy to show that for $\gamma>2/3$ there are no 
solitary-wave solutions. Thus the NPSE gives the condition 
\beq 
-{2\over 3} < {Na_s\over a_{\bot}} < 0  \; , 
\eeq 
for the existence of the 3D Bose-condensed bright soliton under 
transverse confinement. Note however that for $\gamma_c=2/3$ 
the transverse size of the Bose-condensed soliton is not zero, 
in fact $\gamma_c \Phi^2 <1/2$. We have numerically 
verified by solving the 3D GPSE that the critical value $\gamma_c=2/3$ 
is very accurate. This is a remarkable result because,  
contrary to the 3D bright soliton (Eq. 3), 
the widely studied 1D bright soliton (Eq. 5) exists (and it is stable) 
at any $\gamma$: for large values of $\gamma$ 
the wavefunction simply becomes narrower. 
\par 
In order to analyze the stability of our 3D soliton solution and 
its collective modes, it is useful to provide a simple analytical 
representation of its shape. The most natural choice is, as usual, 
a Gaussian with two variational parameters: 
the transverse width $\sigma$ and the axial width $\eta$ [11]. 
In this case, the energy per particle $E$ of the 
condensed atomic-cloud obtained from the 3D GPE energy functional 
is given by 
\beq 
E={1\over 2}\left( {1\over \sigma^2} + {1\over 2 \eta^2} 
+ \sigma^2 - {g \over \sigma^2 \eta} \right)  \; ,   
\eeq 
where $g=\sqrt{2/\pi} \gamma$, 
lengths are in units $a_{\bot}$ and the energy is in units 
$\hbar \omega_{\bot}$. 

\begin{figure}
\centerline{\psfig{file=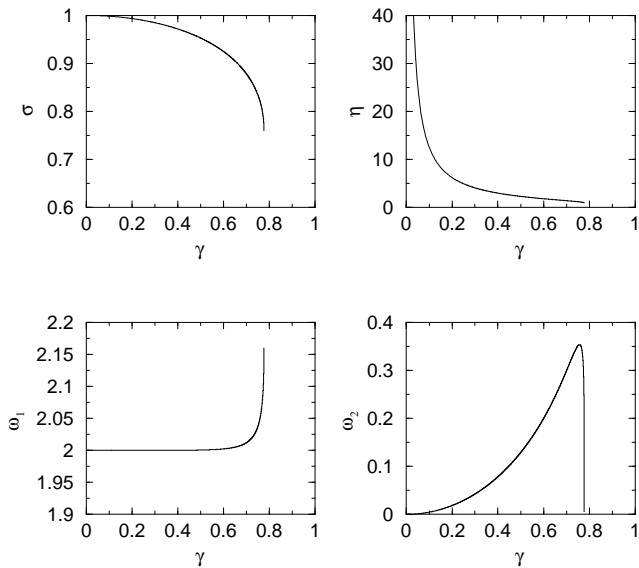,height=3.in}}
\caption{Gaussian approximation of the 3D bright soliton. 
Top: Widths $\sigma$ and $\eta$ as a function of the interaction strength 
$\gamma$. Bottom: Collective frequencies $\omega_1$ and $\omega_2$ 
as a function of the interaction strength $\gamma$, 
with $\gamma=N|a_s|/a_{\bot}$. 
Length in units $a_{\bot}=(\hbar /m\omega_{\bot})^{1/2}$, 
time in units $\omega_{\bot}^{-1}$.} 
\end{figure}

The energy function $E=E(\sigma,\eta)$ has a local minimum, 
that is the 3D bright soliton condition, 
if the inter-atomic strength is smaller than a critical value: 
for larger values the soliton collapses. 
In Figure 2 we plot the widths $\sigma$ and 
$\eta$ of the approximated 3D bright soliton  
as a function of the inter-atomic strength. 
Note that the transverse width $\sigma$ does not change very much 
while the longitudinal width $\eta$ is divergent for 
$\gamma =0$ and it is comparable with $\sigma$ for $\gamma$ 
close to $\gamma_c$. Moreover the critical value $\gamma_c$ of 
the collapse is slightly overestimated with respect to 
the analytical prediction $\gamma_c=2/3$ of NPSE. 
\par 
The Gaussian approximation of the BEC wavefunction 
can be used to study the dynamical stability of the 3D 
bright soliton by means of its collective oscillations. 
The diagonalization of the 
Hessian matrix of the energy function $E(\sigma ,\eta)$ 
gives two frequencies $\omega_1$ 
and $\omega_2$ of collective excitations around the bright soliton solution. 
In Figure 2 we plot such frequencies and the widths $\sigma$ and $\eta$ 
of the solitonic configuration as a function of the strength $\gamma$. 
The frequencies are real and it means that the bright soliton is 
dynamically stable until it collapses at $\gamma_c$. 
Note that only for $\gamma=0$ the frequencies 
$\omega_1$ and $\omega_2$ can be interpreted as transverse and axial  
collective oscillations of the bright soliton; nevertheless 
the mixing angle remains quite small also for finite values of $\gamma$ 
so they can be associated the transverse and axial motion 
respectively. The transverse frequency $\omega_1$ is practically 
constant but close to $\gamma_c$ it suddenly grows. 
The axial frequency $\omega_2$ is zero 
for $\gamma=0$ it increases with $\gamma$ but close to $\gamma_c$ 
it goes to zero. 

\section{Multi-soliton Bose condensates in a box} 

The solitary bright-soliton solution (Eq. 3) has been found 
by using Eq. 1 and Eq. 2. From these equations 
one obtains the Newtonian second-order differential equation 
$$ 
\left[ {d^2\over d\zeta^2} - 2 \gamma 
{\Phi^2\over \sqrt{1-2\gamma\Phi^2} } \right. 
$$ 
\beq
\left. 
+ {1\over 2} 
\left( {1\over \sqrt{1-2 \gamma \Phi^2} } 
+ \sqrt{1-2\gamma \Phi^2} \right) \right] \Phi 
= \mu \Phi \; ,  
\eeq 
where $\zeta =z-vt$ and $\gamma=|a_s|N/a_{\bot}$. 
The constant of motion of this equation is given by  
\beq 
E={1\over 2}\left({d\Phi\over d\zeta} \right)^2 
+ \mu \Phi^2 -\Phi^2\sqrt{1-2\gamma\Phi^2} \; . 
\eeq  
By imposing the boundary condition $\Phi\to 0$ for 
$\zeta \to \infty$, which implies that $E=0$, 
one has the solitary bright-soliton solution, 
that has only one peak.  

\begin{figure}
\centerline{\psfig{file=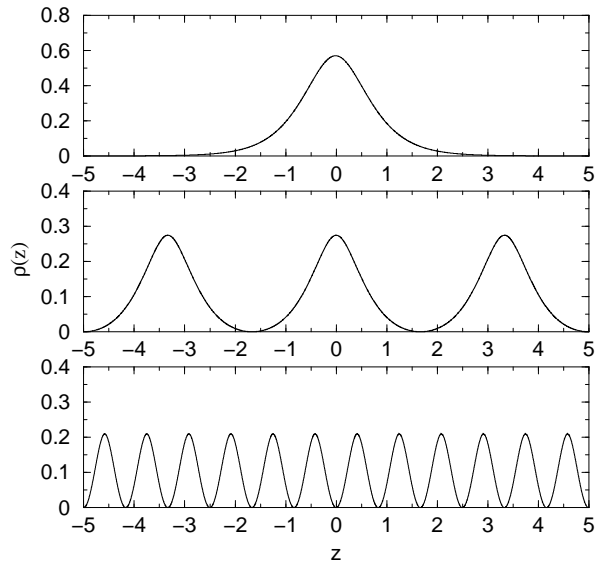,height=3.in}}
\caption{
Axial density profile $\rho(\zeta)$ of the 
bright multi-soliton Bose-condensate in a Box of length $L=10$. 
From top to bottom: one, three and twelve solitons 
close to their critical stength. Units as in Fig. 1.} 
\end{figure}

\par 
It is important to observe that it is possible to obtain 
a multi-soliton static solution, i.e. a multi-peak solution, 
by imposing box boundary conditions: $\Phi(-L/2)=\Phi(L/2)=0$, 
where $L$ is the length of the axial box. 
The density profile is periodic and the quantized number $N_s$ 
of peaks defines the number of solitons in the box for a given 
chemical potential $\mu$ and inter-atomic strength $\gamma$. 
In Figure 3 we plot the axial density profile obtained 
from the numerical integration of Eq. 9 for one, 
three and twelve bright solitons in a box of length $L=10a_{\bot}$. 
Also in the case of 3D bright multi-soliton Bose condensates, 
due to the finite transverse confinement, 
the solution exists only 
if the interaction strength $\gamma=N|a_s|/a_{\bot}$ is 
below a critical value. From the normalization condition 
it is easy to find that the condition of existence for 
a multi-soliton solution is given by 
\beq  
-{L\over 4} < {Na_s} < 0  \; , 
\eeq 
in the limit $N_s\to \infty$. Thus, in a large box, the strength 
$\gamma$ has the critical value $\gamma_c=2/3$ 
for one soliton and $\gamma_c=L/(4a_{\bot})$ for many-solitons. 

\begin{center}
\begin{tabular}{|c|c|c|c|} \hline 
\hspace{0.5cm} $N_s$      \hspace{0.5cm} &
\hspace{0.5cm} $\mu$      \hspace{0.5cm} & 
\hspace{0.5cm} $\rho_0$   \hspace{0.5cm} &
\hspace{0.5cm} $\gamma_c$ \hspace{0.5cm} \\ 
\hline 
    1 & 0.4918 &  0.5694 &  0.6658 \\
    2 & 0.4449 &  0.3311 &  1.2481 \\
    3 & 0.4735 &  0.2745 &  1.6266 \\
    4 & 0.6505 &  0.2491 &  1.8588 \\
    5 & 0.9627 &  0.2353 &  2.0077 \\
    6 & 1.3827 &  0.2275 &  2.1083 \\
    7 & 1.8957 &  0.2231 &  2.1789 \\
    8 & 2.5987 &  0.2178 &  2.2319 \\
    9 & 3.3157 &  0.2159 &  2.2722 \\
   10 & 4.2019 &  0.2134 &  2.3032 \\
   11 & 5.2023 &  0.2113 &  2.3278 \\
   12 & 6.3053 &  0.2098 &  2.3473 \\
   13 & 7.4997 &  0.2085 &  2.3645 \\
   14 & 8.7045 &  0.2082 &  2.3793 \\
   15 & 10.1493 &  0.2069 &  2.3907 \\
   16 & 11.5817 &  0.2068 &  2.4006 \\
   17 & 13.2062 &  0.2061 &  2.4086 \\
   18 & 14.8693 &  0.2055 &  2.4173 \\
   19 & 16.6409 &  0.2051 &  2.4237 \\
   20 & 17.6887 &  0.2050 &  2.4330 \\
\hline 
\end{tabular} 
\end{center}  
{\bf Table 1}. Bose-Einstein condensate with $N_s$ bright 
solitons in a box of length $L = 10 a_{\bot}$. 
Chemical potential $\mu$ and maximum 
density $\rho_0$ of the multi-solitons at 
the critical strength $\gamma_c$. 
Units as in Fig. 1. 
\par 
In Table 1 we show the chemical potential $\mu$, the maximum 
density $\rho_0$ and the critical strength $\gamma_c$ of 
a Bose condensate made by $N_s$ bright solitons 
in a box of length $L = 10 a_{\bot}$. 
As expected the critical strength $\gamma_c$ grows with $N_s$ and 
for large $N_s$ it approaches $L/(4a_{\bot})=2.5$. 
In the case $N_s$ even, the multi-soliton solutions are also solutions 
with periodic boundary conditions (toroidal configuration with 
axial length $L$). In a torus, this train of solitons can travel without 
spreading with arbitrary velocity. 

\section{Propagation of a soliton train: 
comparision with experiments} 

In Ref. [5] it has been reported the formation of a 
multi-soliton Bose condensate of $^7$Li atoms created in a quasi-1D 
optical trap that can be modelled by an asymmetric harmonic 
potential with a large aspect ratio 
($\lambda=\omega_{\bot}/\omega_z=100$). 
In this experiment, a "soliton train", containing 
$N_s$ solitons ($N_s=4,5,6$), has been created from a 
Bose-Einstein condensate by magnetically tuning the 
inter-atomic interaction from repulsive ($a_s=10 a_B$, 
with $a_B$ the Bohr radius) to attractive ($a_s=-3 a_B$) 
using a Feshbach resonance. 
The soliton train has been set in motion and 
observed to propagate in the axial optical harmonic potential 
for many oscillatory cycles with a small spreading. 
Moreover, it has been found that the spacing between solitons is 
compressed at the turning points and spread out at the center of 
the oscillation, suggesting a repulsive interaction between 
neighboring solitons. 

\begin{figure}
\centerline{\psfig{file=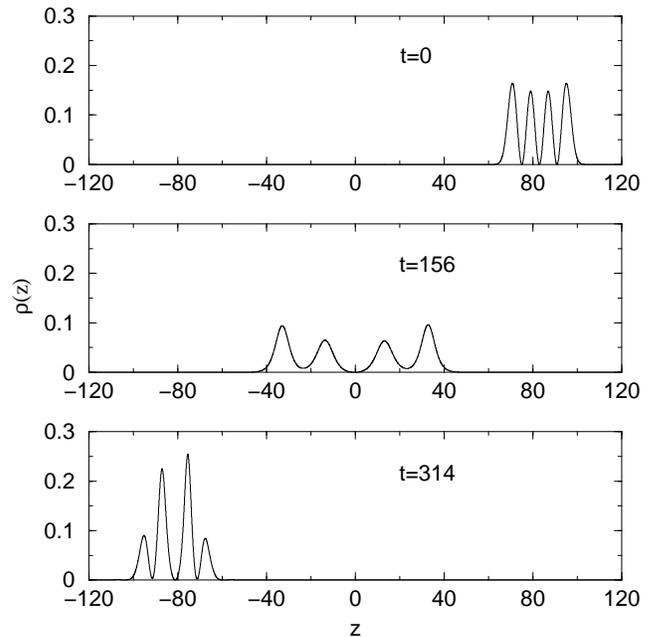,height=3.4in}}
\caption{Axial density profile of the travelling "soliton 
train" of $^7$Li atoms in a anysotropic harmonic potential 
with $\omega_{\bot}=400$ Hz and $\omega_z=4$ Hz. 
Scattering length: $a_s=-3a_B$ with $a_B$ Bohr radius. 
$5000$ atoms for each soliton. Units as in Fig. 1 and Fig. 2.} 
\end{figure}

\par 
By using our time-dependent NPSE it is quite easy to 
simulate the dynamics of the soliton train described in [5]. 
In Figure 4 we plot three frames of the time evolution 
of a set of four $^7$Li bright solitons, each of them made of 
$5000$ atoms, initially located far from the center of the 
axial harmonic potential with an alternating phase structure. 
Note that the train configuration ceases to
be a true solitonic solution as soon as we switch off the axial
confinement. In fact its shape and density change during the motion. 
The agreement between our numerical results of Figure 4 
and the experimental data shown in Figure 4 of Ref. [5] is remarkable: 
the spacing between solitons increases near the center of 
oscillation and bunches at the end points. 
An alternating phase structure is essential to have a repulsion 
between neighboring solitons [12]. In fact, we have verified that 
a train of solitons with the same phase (attractive interaction 
between solitions) becomes a narrow blob of high density 
which can induce the collapse of the condensate. 
In the case of positive scattering length our calculations 
show that during the motion an initially confined Bose condensate 
made of many peaks becomes a blob which spreads along 
the axial trap but partially recomposes at the turning points. 
\par 
The effective interaction between two
bosonic matter waves does indeed depend 
on their phase difference $\Delta$ , being attractive for  
$\Delta > 0$ and repulsive otherwise. 
This phenomenon is fully analogous to what has been already studied in 
the framework of non linear optics and may be simply
understood on the basis of an elementary argument regarding
the extent of the overlap between the two waves. 
From the one dimensional GPE, the interaction energy 
density of the condensate is 
proportional to the square of the local particle density: 
$W(z)\propto a_s \rho(z)^2$. Now we consider a wave function
accurately approximated by the superposition of two well separated 
solitons $\psi(z)
\sim \left [ \phi(z-z_0)e^{i\Delta/2} + \phi(z+z_0)e^{-i\Delta/2}
\right ]/\sqrt{2}$, where $2z_0$ is the relative separation.
By substituting this form into the expression of 
the interaction energy density 
at $z=0$ we get $W_2(0)\propto a_s (\cos\Delta +1)$. A measure 
of the effective interaction can be obtained by considering the
difference between $W_2(0)$ and the analogous self interaction
density of a single soliton $W_1(0)$ which gives: 
$W_2(0)-W_1(0)\propto  a_s\cos\Delta$. In the case of negative
scattering length the interaction turns out to be attractive as
soon as $\cos\Delta >0$ and repulsive otherwise, in agreement with 
more quantitative analysis [12] and previous numerical findings. 

\section{Tunneling with bright solitons} 

Now we investigate the role of solitonic configurations and 
the effect of an attractive inter-atomic interaction 
on the behavior of a Bose-Einstein condensate during tunneling. 
The initial condition of the BEC is the ground-state 
of NPSE with a harmonic trapping potential 
also in the horizontal axial direction: 
$V(z)= {1\over 2}m\omega_z^2 (z-z_0)^2$. 
To have a cigar-shaped condensate 
we choose $\lambda=\omega_{\bot}/\omega_z=10$. 
We set $z_0=20$, where $z_0$ is written 
in units of the harmonic length $a_z=(\hbar/m\omega_z)^{1/2}$. 
For $t>0$ the trap in the axial direction is switched off 
and a Gaussian energy barrier is inserted at $z=0$. The 
potential barrier is given by 
\beq
V(z)= V_0 \; e^{-z^2/ \Sigma^2} \; , 
\eeq 
where $V_0$ is the height of the potential barrier 
and $\Sigma$ its width. 
The BEC is moved towards the barrier 
by adding an initial momentum $p_0$ in the axial direction: 
\beq 
f(z,0) \to f(z,0) \; e^{- i p_0 z/\hbar} \; .  
\eeq 
  
\begin{figure}
\centerline{\psfig{file=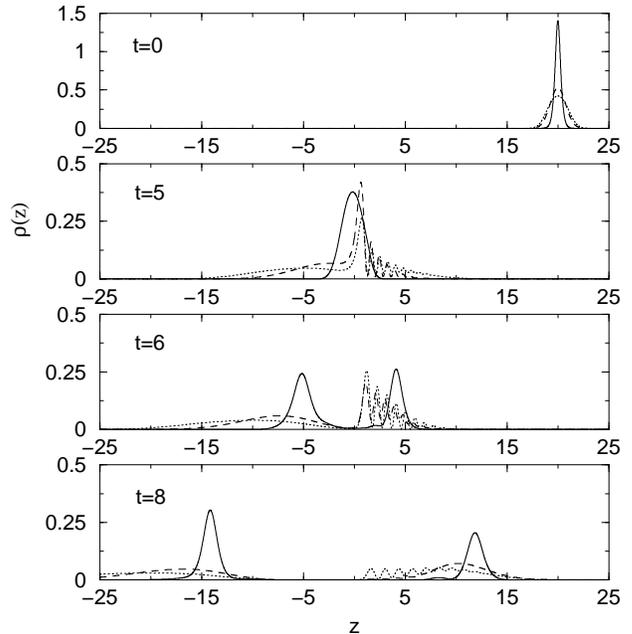,height=3.4in}}
\caption{Axial density profile of the Bose  
condensate tunneling through the Gaussian barrier. 
Start-up kinetic energy per particle 
of the condensate: $E_0=p_0^2/(2m)=10$. 
Gaussian barrier parameters: $V_0=10$ and $\Sigma=1$. 
Three values of the interaction strength: 
$Na_s/a_z=-2\cdot 10^{-1}$ (solid line),  
$Na_s/a_z=0$ (dashed line), 
$Na_s/a_z=2\cdot 10^{-1}$ (dotted line). 
Length in units $a_z=(\hbar /m\omega_z)^{1/2}$, 
time in units $\omega_z^{-1}$, and 
energy in units $\hbar \omega_z$.} 
\end{figure}

We have verified that the condensate with 
a scattering length close to the 
collapse value ($Na_s/a_z=-0.2$) is a bright soliton: 
its shape does not depend on the axial harmonic 
trapping potential. 
In Figure 5 we plot the axial density profile of the Bose 
condensate at different instants for three values of the 
interaction strength: $Na_s/a_z=0.2$, $Na_s/az=0$ and 
$Na_s/a_z=-0.2$. While the case of positive scattering 
length is quite similar to that of zero scattering length, 
the case with negative scattering length is peculiar: 
as previously stated the condensate is a soliton, moreover 
it does not show interference 
patterns and simply splits into a reflected 
and a transmitted matter wave. However, the reflected and the 
transmitted waves are not solitonic, i.e. the tunneling 
through a barrier breaks the solitonic nature of the matter wave.  
\par 
If we decrease further the scattering 
length ($a_s<0$), the Bose condensate becomes unstable. 
For $Na_s/a_z=-0.2$ the Bose-condensed bright soliton  
is still stable but we can induce its collapse 
at the impact time by increasing the initial momentum 
$p_0$ or the energy barrier $V_0$. In fact, 
in these cases the local density of the condensate 
becomes sufficiently large to give rise to the collapse  
of the condensate. An inspection of Eq. 1 shows that a sufficient 
condition for the collapse of the condensate is that 
\beq 
a_s N |f|^2 < - {1\over 2}  \; , 
\eeq 
that is the condition for which condensate shrinks to zero 
in the transverse direction. 
Our numerical computations show that this condition 
gives precisely the one-dimensional density at which there is the 
collapse of the condensate during tunneling: $\rho_{1D}=1/(2|a_s|)$.  

\section{Collisions with bright solitons} 

Solitons show peculiar properties not only in the collision with a 
barrier but also in the collision with other waves. 
In Figure 6 we plot some frames of the collision of 
two Bose-condensed bright solitons obtained by numerically 
solving the time-dependent NPSE (Eq. 1). 

\begin{figure}
\centerline{\psfig{file=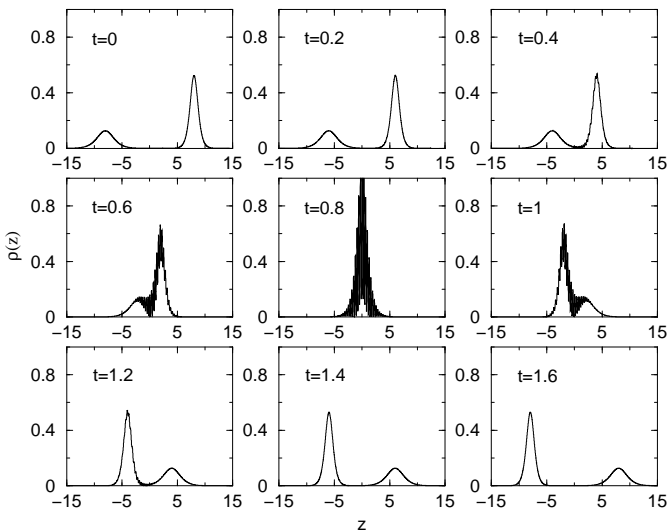,height=2.8in}}
\caption{Collision of two bright solitions with a different number 
of particles: $N_1=2N_2$ and $N_1|a_s|/a_z=0.1$. 
Start-up momentum: $p_0=50$. Units as in Fig. 5.} 
\end{figure}

The two solitons have a different number of particles and 
start with opposite momenta of equal modulus $p_0$. 
As shown in Figure 6, at the impact many fringes of interference 
are produced. We have verified that the number of fringes grows 
with the momentum $p_0$ while the maximal density of the 
interference peak remains constant. 
Moreover, we have found that the only effect of 
a phase-difference between the two colliding 
solitons is a shift in the position of fringes. 
In particular, at a fixed time, by changing 
the phase difference by $\pi$ the spatial locations of maxima 
of interference are shifted into the positions of minima. 
After the impact the two bright solitons continue 
their motion without any recollection of 
the impact: the solitons acquire again their initial shape and then 
continue the motion without shape deformations. 
\par
This phenomenon of transparency can be also seen in the 
collision between a Bose-condensed bright soliton and 
a non-solitonic Bose condensate while two colliding generic matter-waves 
are not transparent. The numerical solution 
of NPSE confirms that, apart the interference at the impact, 
the solitonic matter-wave and the non-solitonic matter-wave do not see 
each other. At large times the bright soliton moves 
with its initial density profile while the non-solitonic wave 
evolves with a spreading that is not influenced by the collision 
with the bright soliton. 
\par 
In the collision process of two Bose-condensed bright solitons 
the maximal density of the interference peak depends on 
the interaction strength, i.e. 
on the initial density of the two colliding solitons. 
For two colliding bright 
solitons with the same number of particles the maximal density 
at impact is about four times 
the initial density of each soliton. We have verified that when the 
maximal density satisfies the condition of Eq. 13 then 
the collapse of the condensates at the impact sets in. 

\section*{Conclusions} 

Our calculations suggest that, contrary to the 1D bright soliton, 
the 3D bright soliton under transverse confinement exists and it 
is dynamically stable only if the attractive inter-atomic 
interaction is smaller than a critical value. We have analytically 
determined this critical interaction strength and the density profile 
of the 3D bright soliton. We have also found the collective oscillations 
of the bright soliton as a function of the interaction strength 
by using a Gaussian approximation. Then we have investigated 
multi-soliton Bose condensates with box boundary conditions. 
Also these multi-soliton solutions exist if the interaction strength 
is below a critical threshold that grows with box length. 
The dynamical properties of a train of solitons have been investigated 
by numerically solving our effective 1D nonpolynomial 
nonlinear Schr\"odinger equation: the results are in qualitative 
agreement with the experimental data. In the case of soliton scattering, 
our theoretical results show that the collapse of 3D soliton can be induced 
by the scattering on a Gaussian barrier when 
the density at the impact reaches a critical value, 
which does not depend on the solitonic nature of the incident wave. 
A clear signature of solitonic behavior is the transparency 
during collisions. Analyzing the collision between bright solitons 
we have verified that, apart the interference at the impact, 
the Bose-condensed soliton is transparent, namely at large times 
it recovers its initial density profile 
and then travels without spreading. However, if the density of the 
interference peak exceeds a critical density the system collapses. 
Moveover we have shown that there is not transparency during tunneling: 
a bright soliton tunneling through a barrier splits into reflected and 
transmitted wave packets which are not solitonic. 

\begin{description}

\item{\ [1]} M.H. Anderson {\it et al.}, Science {\bf 269}, 189 (1995); 
K.B. Davis {\it et al.}, Phys. Rev. Lett. {\bf 75}, 3969 (1995); 
C.C. Bradley {\it et al.}, Phys. Rev. Lett. {\bf 75}, 1687 (1995). 

\item{\ [2]} {\it Bose-Einstein Condensation in Atomic Gases}, vol. 140 
International School of Physics Enrico Fermi, Eds. M. Inguscio, 
S. Stringari, C. Wieman (IOS Press, Amsterdam, 1999). 

\item{\ [3]} P.G. Drazin and R.S. Johnson, {\it Solitons: An Introduction} 
(Cambridge University Press, Cambridge, 1988).

\item{\ [4]} S. Burger {et al.}, Phys. Rev. lett. {\bf 83}, 5198 (1999);   
J. Denshlag {\it et al.}, Science {\bf 287} {\bf 97} (2000). 

\item{\ [5]} K.E. Strecker {\it et al.}, Nature {\bf 417}, 150 (2002). 

\item{\ [6]} W.P. Reinhardt and C.W. Clark, J. Phys. B {\bf 30}, 
L785 (1997); V.M. Perez-Garcia, H. Michinel, and H. Herrero, 
Phys. Rev. A {\bf 57}, 3837 (1998); Th. Busch and J. R. Anglin, 
Phys. Rev. Lett. {\bf 87} 010401 (2001). 

\item{\ [7]} E.P. Gross, Nuovo Cimento {\bf 20}, 454 (1961); 
L.P. Pitaevskii, Zh. Eksp. Teor. Fiz. {\bf 40}, 
646 (1961) [English Transl. Sov. Phys. JETP {\bf 13}, 451 (1961)]. 

\item{\ [8]} L. Salasnich, A. Parola, and L. Reatto, 
Phys. Rev. A {\bf 65}, 043614 (2002). 

\item{\ [9]} L. Salasnich, Laser Physics {\bf 14}, 198 (2002). 

\item{\ [10]} L. Salasnich, A. Parola, and L. Reatto, 
Phys. Rev. A {\bf 64}, 023601 (2001). 

\item{\ [11]} L. Salasnich, Int. J. Mod. Phys. B {\bf 14}, 1 (2000). 

\item{\ [12]} J.P. Gordon, Opt. Lett. {\bf 8}, 596 (1983). 

\end{description}
              
\end{document}